\documentclass[11pt,a4paper]{article}

\usepackage{mathpple}
\usepackage{graphicx}
\usepackage{url}
\usepackage{paralist}

\usepackage{geometry}
\usepackage{fancyhdr}
\begin{document}

\thispagestyle{plain}

\noindent \textbf{Preprint of:}\\
D. K. Gramotnev and T. A. Nieminen\\
``Rigorous analysis of grazing-angle scattering of
  electromagnetic waves in periodic gratings''\\
\textit{Optics Communication} \textbf{219}, 33--48 (2003)

\hrulefill

\begin{center}

\textbf{\LARGE
Rigorous analysis of grazing-angle scattering of
  electromagnetic waves in periodic gratings
}

{\Large
D. K. Gramotnev$^1$ and T. A. Nieminen$^2$}

{}$^1$Applied Optics Program, School of Physical and Chemical Sciences,
Queensland University of Technology, GPO Box 2434, Brisbane,
QLD 4001, Australia; e-mail: d.gramotnev@qut.edu.au

{}$^2$Centre for Biophotonics and Laser Science, Department of Physics,
The University of Queensland, Brisbane, Qld 4072, Australia

\vspace{1cm}

\begin{minipage}{0.8\columnwidth}
\section*{Abstract}
Grazing-angle scattering (GAS) is a type of Bragg scattering of waves
in slanted non-uniform periodic gratings,
when the diffracted order satisfying the Bragg condition
propagates at a grazing angle with respect to the boundaries of
a slab-like grating. Rigorous analysis of GAS of bulk TE
electromagnetic waves is undertaken in holographic gratings
by means of the enhanced T-matrix algorithm. A comparison of
the rigorous and the previously developed approximate
theories of GAS is carried out. A complex pattern of numerous
previously unknown resonances is discovered and
analysed in detail for gratings with large amplitude, for
which the approximate theory fails. These resonances are
associated not only with the geometry of GAS, but are also
typical for wide transmitting gratings. Their dependence on
grating amplitude, angles of incidence and scattering, and
grating width is investigated numerically. Physical interpretation
of the predicted resonances is linked to the existence and the
resonant generation of special new eigenmodes of
slanted gratings. Main properties of these modes and their
field structure are discussed.
\end{minipage}

\end{center}

\section{Introduction}

Grazing-angle scattering (GAS) is a strongly
resonant type of wave scattering in uniform, strip-like,
slanted, wide periodic gratings [1]. It is realised
when the scattered wave (the first diffracted
order) propagates almost parallel to the front
grating boundary, i.e. at a grazing angle with
respect to this boundary. Thus, GAS is intermediate
between extremely asymmetrical scattering (EAS)
(that occurs when the scattered wave propagates
parallel to the grating boundaries [2­6]) and
conventional Bragg scattering in reflecting or
transmitting gratings (where the scattered wave
propagates at a significant angle with respect to
the grating boundaries).

   GAS is characterised by a unique combination
of two simultaneous resonances [1]. One of these
resonances is with respect to frequency of the
incident wave. This resonance is typical for EAS
and is related to a strong increase in the scattered
wave amplitude inside and outside the grating at a
resonant frequency [1--6]. The other resonance is
with respect to angle of scattering (we will call it
GAS resonance) [1]. It occurs at a grazing resonant
angle of scattering when the scattered wave (the $+1$
diffracted order) propagates almost parallel to the
front boundary into the grating [1]. The GAS
resonance may result in further strong increase of
the scattered wave amplitude inside (especially in
the middle of) the grating [1]. It is important that
in the GAS resonance, not only the scattered wave
amplitude experiences a strong resonant increase
compared to its values typical for EAS, but also
the amplitude of the incident wave resonantly
increases in the middle of the grating [1]. The larger
the grating width and/or the grating amplitude, the
stronger and sharper the GAS resonance for both
incident and scattered waves [1]. Note that this is
the complete opposite to the general tendency for
EAS, where increasing grating amplitude and/or
grating width results in decreasing scattered wave
amplitude [3--9].

    It has been demonstrated that one of the main
physical reasons for the unique wave behaviour
during GAS is the diffractional divergence of the
scattered wave (that is similar to divergence of a
laser beam of finite aperture). The necessity of
taking the divergence into account can easily be
seen in the geometry of EAS, i.e. when the scattered
wave propagates parallel to the grating
boundaries. Indeed, scattering occurs only within
the grating, and the scattered wave propagates
parallel to the grating. Therefore, if the diffractional
divergence is ignored, the scattered beam
must be confined to the grating region, i.e. must
have an aperture that is equal to the grating width.
Obviously such a beam experiences diffractional
divergence into the regions outside the grating.
For more detailed discussion of the role of
diffractional divergence in EAS see [3--6]. Since GAS
is intermediate between EAS and the conventional
scattering, the diffractional divergence of the
scattered wave plays an important role for it as
well [1]. In particular, it was shown that strong
GAS resonance occurs only in sufficiently wide
gratings of widths that are larger than a critical
width. Physically, half of the critical width is equal
to the distance within which the scattered wave can
be spread across the grating by means of the
diffractional divergence, before being re-scattered by
the grating [7,8].

    On the basis of understanding the role of the
diffractional divergence, a new powerful approximate
method of analysis of GAS and EAS has
been developed [1,3--8]. This approach is directly
applicable for all types of waves, including surface
and guided optical and acoustic waves in wide
periodic groove arrays [1,3,4,6]. Rigorous analysis
of EAS of bulk electromagnetic waves in volume
holographic gratings [9] has confirmed a high
degree of accuracy of the approximate approach in
gratings where the scattered wave amplitude
experiences a strong resonant increase. Simple
applicability conditions for the approximate theory
have been derived [1] and verified by means of the
rigorous analysis [9].

    However, the excellent agreement of the
approximate and rigorous theories for EAS in
gratings with sufficiently small grating amplitude [9]
does not automatically extend to the case of GAS.
This is because the GAS resonance results in a
further substantial increase of amplitudes of the
$+1$ and 0th diffracted orders (scattered and incident
waves) in the grating. Since these diffracted
orders are directly coupled to the $-1$ and $+2$
diffracted orders [9], the amplitudes of the $-1$ and $+2$
orders must also significantly increase even if the
grating amplitude is small (if the scattered wave
amplitude is sufficiently large). As a result, the
approximate theory (neglecting all diffracted
orders other than the zeroth and the first orders) is
expected to fail if the GAS resonance is sufficiently
strong. This is especially the case for larger grating
amplitudes that are associated with larger amplitudes
of higher diffracted orders and stronger GAS
resonance [1]. In addition the analysis of GAS in
gratings with large amplitude may result (and this
is clearly confirmed by this paper) in radically new
physical phenomena in slanted gratings.

   Therefore, the aim of this paper is to present a
rigorous analysis of GAS and associated resonances
in wide holographic gratings with small
and large grating amplitude. The rigorous analysis
of DEAS will be carried out by means of the
enhanced T-matrix algorithm [10,11]. Applicability
conditions for the approximate theory [1] will
be verified, and the comparison between the
approximate and rigorous analyses will be carried
out. The analysis will not only be confined to the
geometry of GAS. It will be shown that scattering
in the conventional transmitting gratings with
large amplitude is also characterised by very
strong GAS-like resonances, the analysis of which
will be carried out in detail. New special eigenmodes
guided by a grating will be shown to exist,
analysed and used for the interpretation of the
predicted resonances.

\section{ Structure and solutions}

\begin{figure}[!b]
\centerline{\includegraphics[width=0.5\columnwidth]{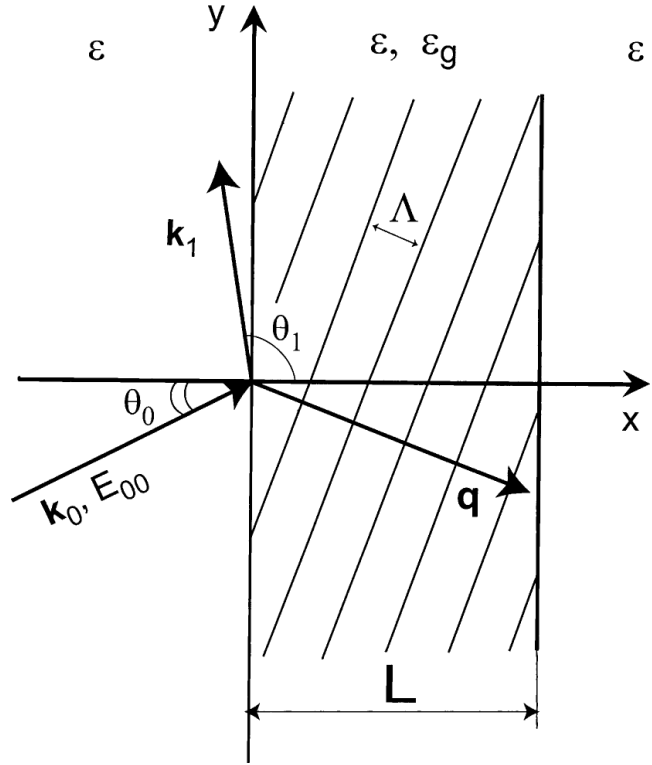}}
\caption{The scheme for grazing-angle scattering of bulk TE
electromagnetic waves in a slanted holographic grating of width $L$
and grating amplitude $\epsilon_g$. The angle of
incidence is $\theta_0$ and the
amplitude of the incident wave in front of the grating is $E_{00}$. The
Bragg condition is satisfied precisely for the first diffracted order
(scattered wave) that propagates at an angle $\theta_1$
that is close to $\pi/2$
(i.e. at a grazing angle with respect to the grating boundaries).}
\end{figure}

   Consider an isotropic medium with a volume,
uniform, holographic grating represented by
sinusoidal variations of the dielectric permittivity
(Fig. 1):
\begin{eqnarray}
\epsilon_s & = & \epsilon + \epsilon_g
\exp(\mathrm{i}q_xx + \mathrm{i}q_yy)  + \epsilon_g^\ast
\exp(-\mathrm{i}q_xx - \mathrm{i}q_yy) \nonumber \\
& & \textrm{if} \,\,\,\, 0<x<L, \nonumber\\
\epsilon_s = \epsilon \,\,\,\, \textrm{if} \,\,\,\, x<0
\,\,\,\, \textrm{or} \,\,\,\, x>L,
\end{eqnarray}
where $L$ is the width and eg is the amplitude of the
grating, the mean dielectric permittivity, $\epsilon$, is the
same throughout the structure (see Eq. (1)), $q_x$ and
$q_y$ are the $x$- and $y$-components of the reciprocal
lattice vector $\mathbf{q}$, $q = 2\pi/\Lambda$,
$\Lambda$ is the period of the
grating; the co-ordinate system is shown in Fig. 1.
It is also assumed that there is no dissipation of
electromagnetic waves inside and outside the
grating ($\epsilon$ is real and positive), and the structure is
infinite along the $y$- and $z$-axes. A bulk TE
electromagnetic plane wave with the amplitude $E_{00}$
and wave vector $\mathrm{k}_0$ is incident onto the grating at
an angle $\theta_0$ in the $xy$-plane---Fig. 1 (non-conical
scattering).

The solutions inside and outside the non-uniform
grating can be written as [9--11]:
\begin{equation}
E_1(x,y) = \sum_{n=-\infty}^{+\infty} S_n(x)
\exp(\mathrm{i}k_{nx}x+\mathrm{i}k_{ny}y),
\end{equation}
\begin{equation}
E|_{x<0} = E_{00}\exp(\mathrm{i}\mathbf{k}_0\cdot\mathbf{r})
+ \sum_{n=-\infty}^{+\infty} A_n
\exp(\mathrm{i}\mathbf{k}_{\mathrm{r}n}\cdot\mathbf{r}),
\end{equation}
\begin{equation}
E|_{x>L} = \sum_{n=-\infty}^{+\infty} B_n
\exp(\mathrm{i}\mathbf{k}_{\mathrm{t}n}\cdot\mathbf{r}
-\mathrm{i}k_{\mathrm{t}nx}L),
\end{equation}
where the dependence of the field on time, i.e. the
factor $\exp(-\mathrm{i}\omega t)$, is omitted,
$k_{nx}$ and $k_{ny}$ are the $x$-
and $y$-components of the wave vectors $\mathbf{k}_n$
determined by the Floquet condition:
\begin{equation}
\mathbf{k}_n = \mathbf{k}_0 - n\mathbf{q} \,\,\,\,
(n=0,\pm 1, \pm 2, ... )
\end{equation}
and the components of the wave vectors $\mathbf{k}_{\mathrm{r}n}$
and $\mathbf{k}_{\mathrm{t}n}$
of the nth reflected and transmitted waves in the
regions $x < 0$ and $x > L$ are determined by the
equations:
\begin{equation}
k_{\mathrm{r}ny} = k_{\mathrm{t}ny}, \,\,\,\,
k_{\mathrm{r}nx} = - k_{\mathrm{t}nx} = - (k_0^2 - k_{ny}^2)^{1/2}.
\end{equation}
If $k_{ny} \le k_0$, then $k_{\mathrm{r}nx} \le 0$ and
$k_{\mathrm{t}nx} \ge 0$ (propagating
waves), while if $k_{ny} > k_0$, then
Im$(k_{\mathrm{r}nx}) < 0$ and Im$(k_{\mathrm{t}nx}) > 0$
(evanescent waves).

   The Bragg condition is assumed to be satisfied
precisely for the $+1$ diffracted order in Eq. (2) for
all angles of scattering h1 that are assumed to be
close to $\pi/2$ (the geometry of GAS---Fig. 1). The
fact that the Bragg condition is satisfied precisely
for all values of $\theta_1$ means that we have to adjust
the grating parameters (the period and slanting
angle) for each value of $\theta_1$. This is inconvenient
in practice, but allows the analysis of GAS at
optimal conditions, i.e. separately from the effects
of small detunings of the Bragg condition (the
same was done in [1] for the approximate theory of
GAS).

   The $x$-dependencies of amplitudes of the
diffracted orders inside the grating are obtained from
the solution of the truncated rigorous coupled
wave equations and the boundary conditions at
the grating boundaries $x = 0, L$. As mentioned
above, the rigorous analysis of the coupled wave
equations and boundary conditions is carried out
by means of the numerically stable enhanced
T-matrix algorithm developed by Moharam et al.
[10,11].

\section{GAS in gratings with small amplitude}

   As indicated in the introduction and paper [1],
GAS is characterised by a strong resonance with
respect to angle of scattering $\theta_1$ (GAS resonance)
only if the grating width $L$ is greater than the
critical width $L_c$. Therefore, in this paper, we will
consider mainly wide gratings with $L > L_c$.
Physically, half of the critical width is equal to the
distance within which the scattered wave can be
spread across the grating by means of the diffractional
divergence, before being re-scattered by the
grating [7,8]. Two simple methods for the
determination of Lc have been described in [7,8].

   For example, consider a volume holographic
grating with the parameters: $\epsilon = 5$,
$\theta_0 = \pi/4$, and
the wavelength in vacuum $\lambda = 1\mu$m. The grating
amplitude eg and grating width will be varied
below (however, we will always have $L > L_c$). The
Bragg condition is assumed to be satisfied precisely
for the $+1$ diffracted order, i.e. in Eq. (5),
$k_1 = \epsilon^{1/2}\omega c$, and $\mathbf{k}_1$
is almost parallel to the
grating boundaries---Fig. 1.

\begin{figure}[!t]
\centerline{\includegraphics[width=0.7\columnwidth]{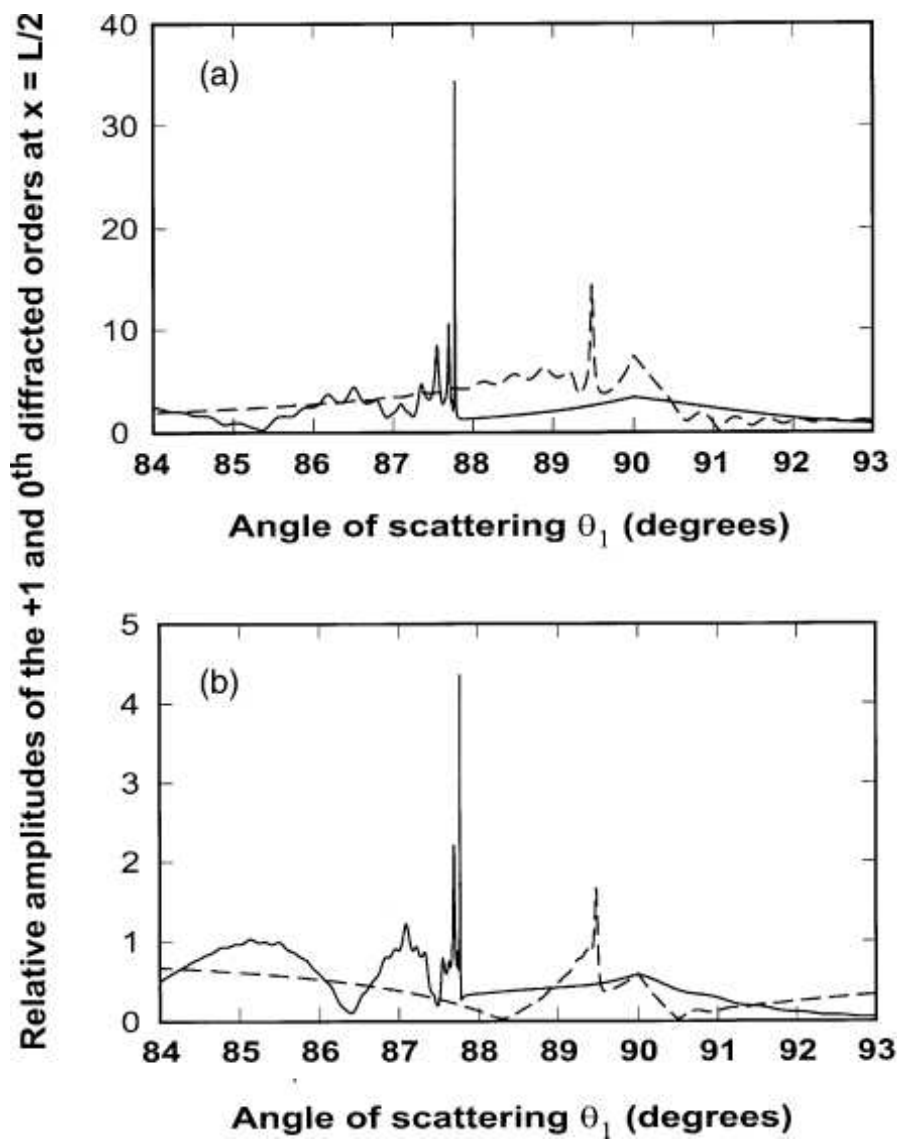}}
\caption{The rigorous dependencies of the relative amplitudes of
(a) the scattered wave (the $+1$ diffracted order),
$|S_1/E_{00}|$, and (b)
the incident wave (the 0th diffracted order),
$|S_0/E_{00}|$, in the
middle of the grating (i.e. at $x = L/2$) on angle of scattering
$\theta_1$.
The structural parameters: $\epsilon = 5$, $\theta_0 = 45^\circ$,
the wavelength $\lambda(\mathrm{vacuum}) = 1\mu$m, $L = 64.5\mu$m,
the period and orientation of
the gratings are determined by the Bragg condition and $\theta_1$.
Dashed curves: $\epsilon_g = 5 \times 10^{-3}$;
solid curves: $\epsilon_g = 5 \times 10^{-2}$.}
\end{figure}

   Typical rigorous dependencies of amplitudes of
the $+1$ and 0th diffracted orders (the scattered and
incident waves) in the middle of the grating (i.e. at
$x = L/2$) on angle of scattering h1 are presented in
Figs. 2(a) and (b) for the gratings of width
$L = 64.5\mu$m, and $\epsilon_g = 5 \times 10^{-3}$ (dashed curves)
and $\epsilon_g = 5 \times 10^{-2}$ (solid curves). The reasons for
choosing $L = 64.5\mu$ will become clear below.

   It can be seen that a very strong resonant increase
   of the scattered (Fig. 2(a)) and incident (Fig.
2(b)) wave amplitudes occurs at a resonant angle
of scattering $\theta_{1\mathrm{r}} < \pi/2$, i.e. when the scattered
wave propagates at a grazing angle into the grating.
It is important that these amplitudes in the
GAS resonance are much larger than those typical
for EAS (i.e. at $\theta_1 = \pi/2$---Figs. 2(a) and (b)).
It can also be seen that the larger the grating
amplitude, the stronger the GAS resonance---compare
solid and dashed curves in Figs. 2(a) and
(b) (the same conclusion as in the approximate
theory [1]). In addition, the resonant angle
noticeably decreases with increasing grating
amplitude (Figs. 2(a) and (b)).

    Similarly to the approximate theory, the
rigorous analysis suggests that the GAS resonance
increases with increasing grating width $L$. It can also
be seen that for given structural parameters there
is a set of approximately periodically spaced
optimal grating widths $L_\mathrm{opt}$ for which the scattered
wave amplitude outside the grating is zero
(amplitudes $A_1$ and $B_1$ in Eqs. (3) and (4) are zero), and
the energy of the scattered wave is entirely
concentrated within the grating region---see also [1].
Recall that since the mean dielectric permittivity is
the same inside and outside the grating (see Eq.
(1)), there is no conventional guiding effect in the
grating. The entire localisation of the scattered
wave inside the grating (at $L = L_\mathrm{opt}$) is achieved
due to peculiarities of scattering, re-scattering and
diffractional divergence (similarly to the approximate
theory [1]). It can also be seen that the values
of Lopt and the corresponding resonant angles of
scattering $(\theta_{1\mathrm{r}})_\mathrm{opt}$
obtained from the rigorous theory at
$\epsilon_g = 5 \times 10^{-3}$ (small grating amplitude) are
only insignificantly different from those obtained
in the approximate theory [1], if the grating width
does not exceed $\approx 100\mu$m. At the same time, for
very large grating widths, or large grating amplitudes,
the rigorous theory gives different values of
$L_\mathrm{opt}$. These values can be found in the rigorous
theory in the same way as it was done in [1]. The
considered grating width $L = 64.5\mu$m (Figs. 2(a)
and (b)) is one of $L_\mathrm{opt}$ if
$\epsilon_g = 5 \times 10^{-3}$ (dashed
curves in Figs. 2(a) and (b)).

   However, it is important to remember that
strong GAS resonance occurs not only at $L = L_\mathrm{opt}$,
but at any other width that is larger than the
critical width (for the considered structures,
$L_c \approx 28\mu$m for $\epsilon_g = 5 \times 10^{-3}$,
and $L_c \approx 6.5\mu$m for $\epsilon_g = 5 \times 10^{-2}$
[7,8])---see solid curves in Figs. 2(a)
and (b). The main effect of optimal grating widths
for small grating amplitudes is the total localisation
of the scattered wave inside the grating. If the
grating amplitude is larger than $\approx 0.1\epsilon$,
then at
$L = L_\mathrm{opt}$, the scattered wave amplitude at the
grating boundaries may be of the order of the
incident wave amplitude in front of the grating (see
below).

\begin{figure}[!htpb]
\centerline{\includegraphics[width=0.7\columnwidth]{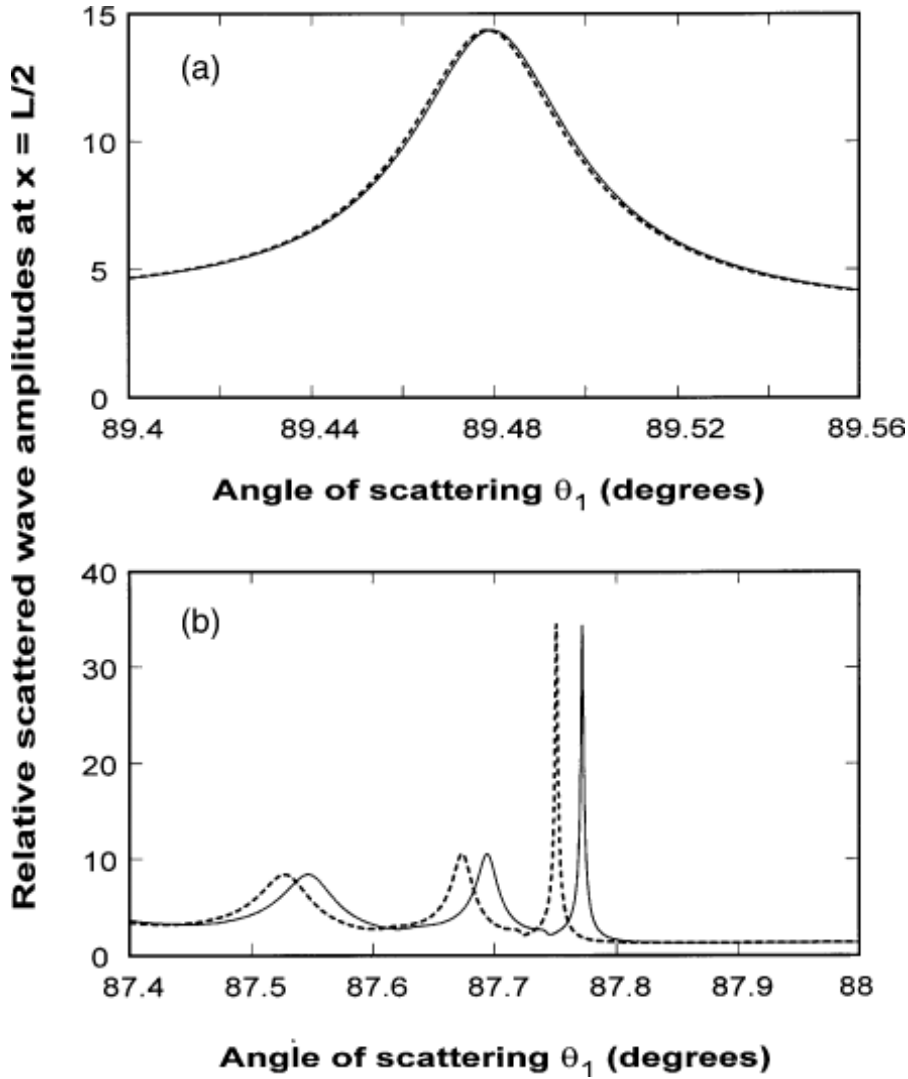}}
\caption{The comparison of the rigorous angular dependencies of
the scattered wave amplitudes in the middle of the grating,
presented in Fig. 2(a) (solid curves), with the same dependencies
obtained from the approximate theory [1] (dotted curves).
(a) $\epsilon_g = 5 \times 10^{-3}$,
(b) $\epsilon_g = 5 \times 10^{-2}$.}
\end{figure}

   If we plot the corresponding approximate
curves (obtained in [1]) in Figs. 2(a) and (b), we
will hardly notice any difference from the
presented rigorous curves. However, this is not
because the approximate and rigorous theories give
the same results, but rather the scale in Figs. 2(a)
and (b) is too small to see the differences between
the curves near sharp resonances. Therefore, Figs.
3(a) and (b) present the rigorous and approximate
angular dependencies of the scattered wave
amplitudes just near the GAS resonance. It can
clearly be seen that if the GAS resonance is not
very strong and the grating amplitude is small,
then the agreement between the approximate and
rigorous theories is very good (within $\approx 2$\%---Fig.
3(a)). At the same time, if the grating amplitude is
increased 10 times, then the GAS resonance
becomes much stronger and sharper, which results
in significant discrepancies between the approximate
and rigorous theories (Fig. 3(b)). Note
however that these discrepancies are mainly limited
to variations of the resonant angle of scattering.
Indeed, as seen from Figs. 3(a) and (b), the
shape and the height of maximums of the rigorous
and approximate dependencies are practically
identical, and the only difference is that the rigorous
curve is slightly shifted to the right (in the
direction of larger angles of scattering). That is,
values of the resonant angle $\theta_{1\mathrm{r}}$ in the rigorous
theory are larger than those predicted by the
approximate theory [1].

   It is interesting that reducing the angle of incidence
$\theta_0$ results in noticeably better agreement
between the approximate and rigorous theories,
despite increasing strength of the GAS resonance
for smaller angles of incidence. For example, if the
angle of incidence $\theta_0 = 0$, then the maximums
similar to those in Fig. 3(b) practically merge, and
the corresponding discrepancies between the
approximate and rigorous theories are no more than
those in Fig. 3(a).

\begin{figure}[htb]
\centerline{\includegraphics[width=0.7\columnwidth]{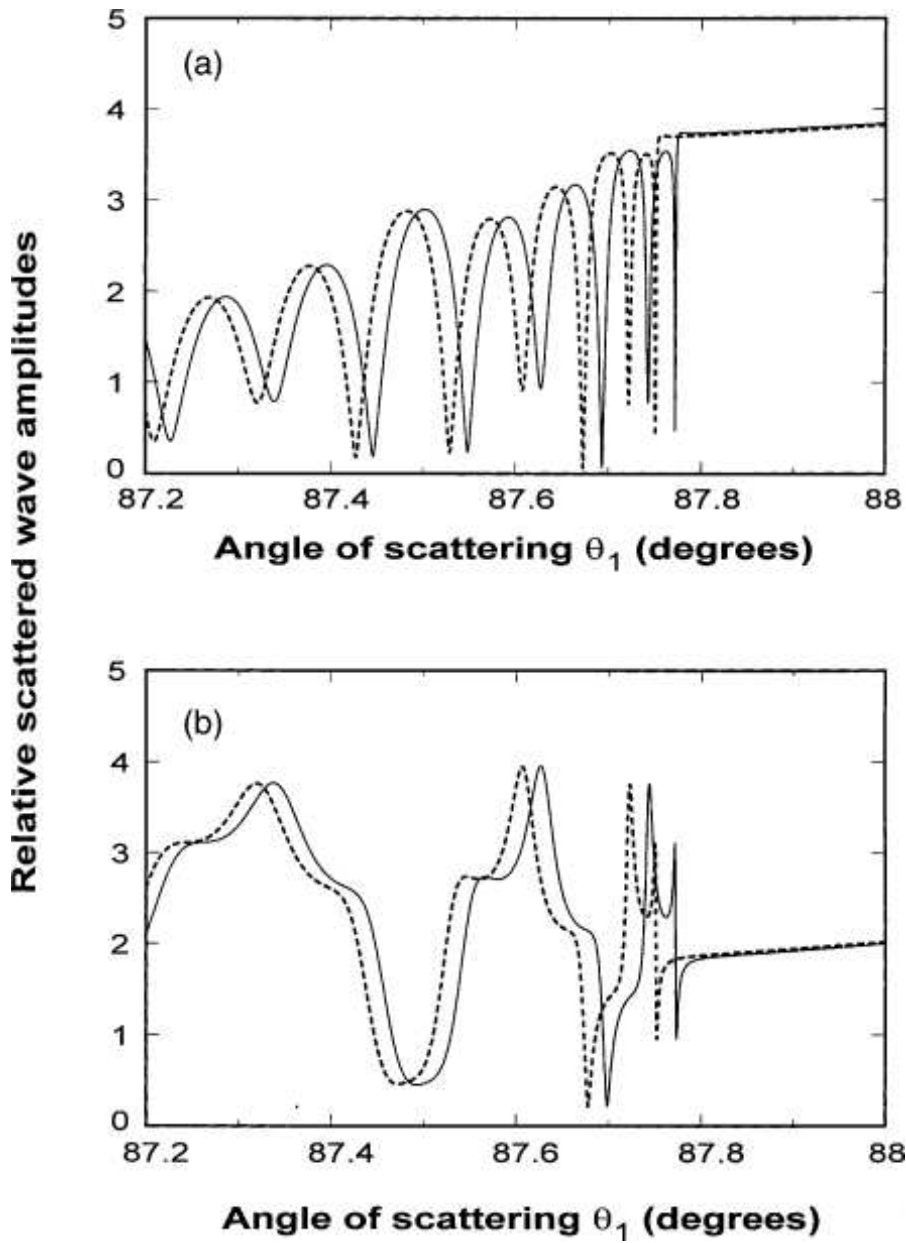}}
\caption{The comparison of the rigorous (solid curves) and
approximate (dotted curves) angular dependencies of the scattered
wave amplitudes at the front (a) and rear (b) grating boundaries
for the structure corresponding to the solid curves in Figs. 2(a)
and (b) (and to Fig. 3(b)), i.e. with $\epsilon_g = 5 \times 10^{-2}$,
$\epsilon = 5$,
$\theta_0 = 45^\circ$, $\lambda(\mathrm{vacuum}) = 1\mu$m,
$L = 64.5\mu$m.}
\end{figure}

    Similar conclusions can be drawn when
considering rigorous angular dependencies of the
scattered wave amplitudes at the front and rear
boundaries of the grating with
$\epsilon_g = 5 \times 10^{-2}$---Figs. 4(a)
and (b). These dependencies are practically
the same as those obtained by means of the
approximate theory [1], except for that they are
slightly shifted towards larger angles of scattering
by the same value as the rigorous curves in Fig.
3(b). Decreasing angle of scattering results in
better agreement between the approximate and
rigorous theories.

    Note that this result is fairly unexpected. One
could think that taking into account higher
diffracted orders in the rigorous theory should result
in significant changes in the pattern of scattering,
especially in the case of strong and sharp oscillations
as in Fig. 4. In particular, these expectations
could be related to additional energy losses from
the grating due to boundary scattering and additional
propagating orders outside the grating [9].
However, as seen from Figs. 2--4, the only effect of
higher diffracted orders in the rigorous theory is a
slight shift of the corresponding dependencies towards
larger angles of scattering. In particular, this
is probably because GAS is characterised by a
strong resonant increase of the incident and scattered
wave amplitudes in the middle of the grating,
while at the boundaries ($x = 0$ and $x = L$) the
scattered wave amplitudes normally do not exceed
a few amplitudes of the incident wave, especially if
the grating amplitude is large. Moreover, if
$L = L_\mathrm{opt}$ and
$\theta_1 = (\theta_{1\mathrm{r}})_\mathrm{opt}$,
then the scattered wave
amplitude at the grating boundaries and outside
the grating is approximately zero. Therefore, the
amplitudes of propagating orders outside the
grating (in particular, the amplitude of boundary
scattered wave) must be small, as well as the
associated energy losses. Thus their effect on the
pattern of scattering is limited.

\section{Scattering in gratings with large amplitude}

\begin{figure}[!t]
\centerline{\includegraphics[width=0.7\columnwidth]{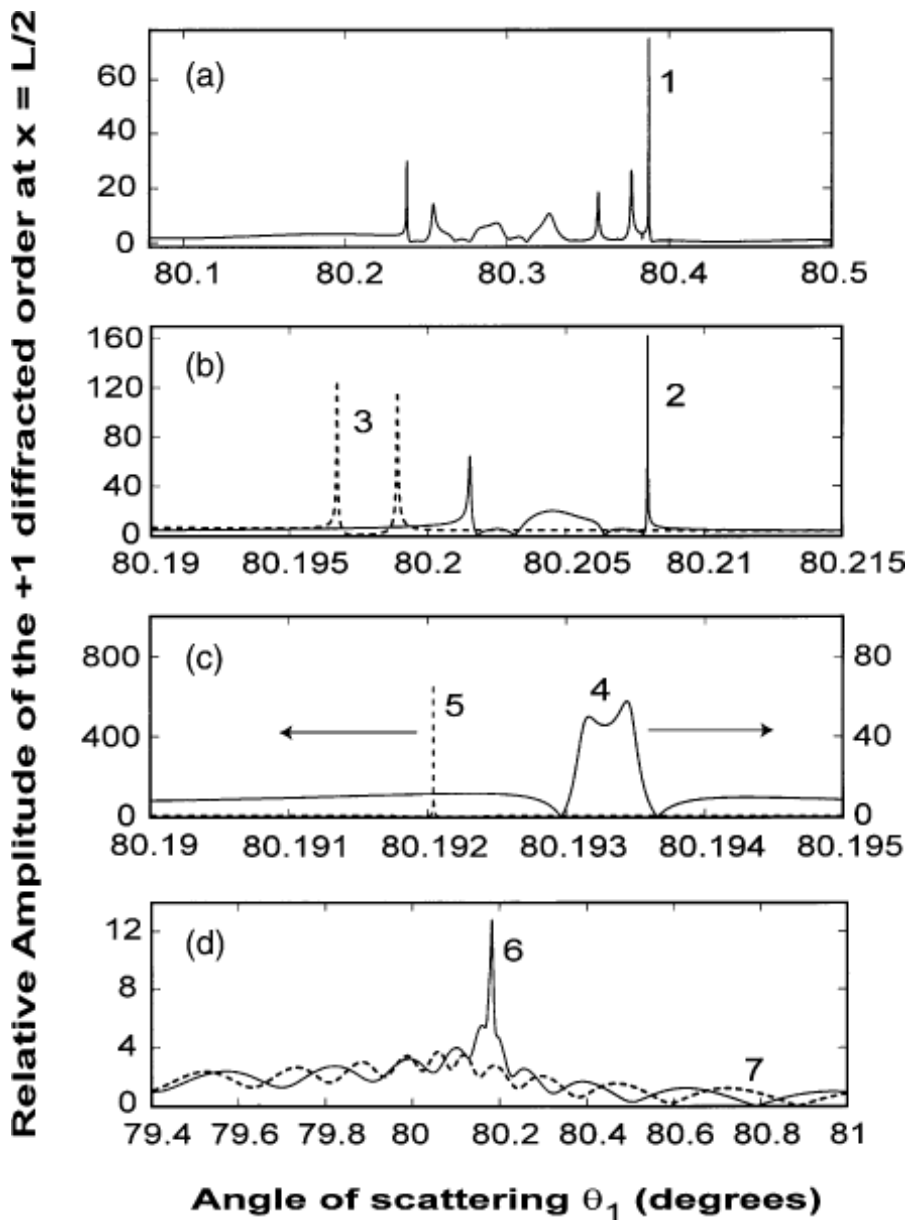}}
\caption{The rigorous dependencies of the relative scattered
wave amplitude $|S_1/E_{00}|$ in the middle of the grating on angle
of scattering $\theta_1$ for large grating amplitudes: (1)
$\epsilon_g = 0.55$, (2) $\epsilon_g = 0.59$,
(3) $\epsilon_g = 0.593$, (4) $\epsilon_g = 0.595$,
(5) $\epsilon_g = \epsilon_{gc1} \approx 0.595562$,
(6) $\epsilon_g = 0.6$, (7) $\epsilon_g = 0.65$.
The other structural parameters are as for Figs. 2--4.}
\end{figure}

   All the results and tendencies presented in the
previous section are true not only for very small
grating amplitudes (Figs. 2--4), but also for values
of $\epsilon_g$ up to $\approx 10$\%
of the mean dielectric permittivity
in the grating (for the considered gratings this is
$\epsilon_g \approx 0.5$). However, when the grating amplitude
approaches $approx 0.1\epsilon$, the rigorous pattern of GAS
changes substantially. For example, if the grating
amplitude $\epsilon_g = 0.55$ and all other parameters are
the same as for Figs. 2--4, then the rigorous
dependencies of the scattered (and incident) wave
amplitude on angle of scattering $\theta_1$ in the middle of
the grating display sharp oscillations between two
distinct resonant peaks (Fig. 5(a)). Outside the
range of angles between these two peaks the scattered
wave amplitude is relatively small and only
weakly depends on $\theta_1$---Fig. 5(a). Increasing
grating amplitude results in shifting these two peaks
towards each other---see Fig. 5(b). In this case both
the peaks become rapidly sharper and higher. This
is especially relevant to the left maximum that may
even overtake the main GAS resonance (i.e. the
right maximum in Fig. 5(a) and (b)). The number of
peaks in between the extreme left maximum and the
extreme right maximum reduces with increasing
grating amplitude. Finally, the two extreme maximums
come so close to each other that they can
hardly be resolved (solid curve in Fig. 5(c)) and then
merge together in one extremely sharp and high
maximum (dotted curve in Fig. 5(c)). The merger
occurs at a critical grating width $\epsilon_{gc1}$
(the reason for
using the index 1 in $\epsilon_{gc1}$ will be clear below). In the
considered case, $\epsilon_{gc1} = 0.595562$---Fig. 5(c).
Further increase of the grating amplitude results in a
rapid decrease of the GAS resonance at
$\epsilon_g = 0.6$---solid curve in Fig. 5(d), with its complete
disappearance at $\epsilon_g = 0.65$----dotted curve in Fig. 5(d).

\begin{figure}[htb]
\centerline{\includegraphics[width=0.7\columnwidth]{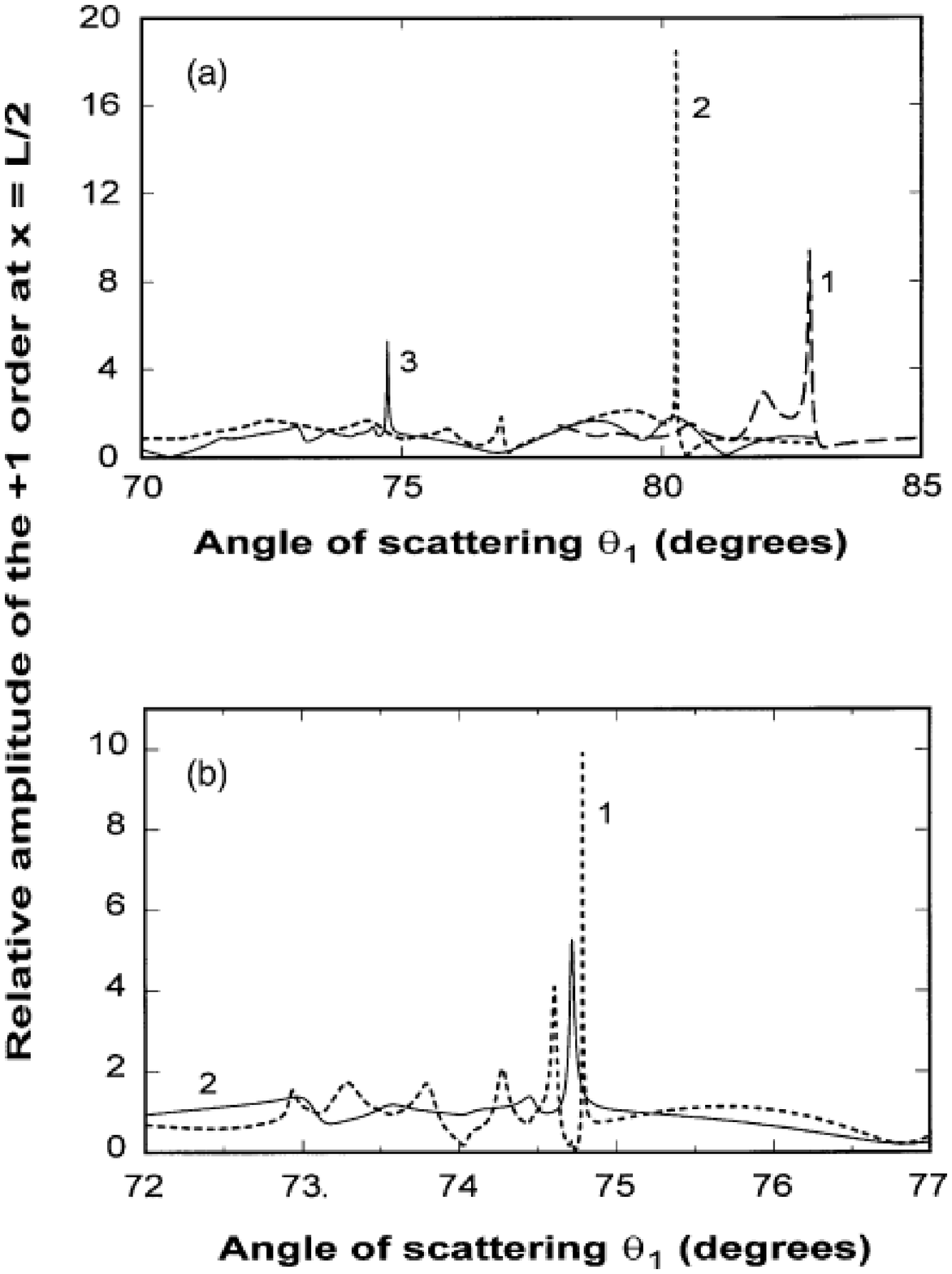}}
\caption{The same dependencies as in Fig. 5, but for different
grating width $L = 20\mu$m, and the following grating amplitudes:
(1) $\epsilon_g = 0.38$, (2) $\epsilon_g = 0.44$,
(3) $\epsilon_g = 0.51$, (4) $\epsilon_g = 0.565$,
(5) $\epsilon_g = 0.58$, (6) $\epsilon_g = 0.5868$.}
\end{figure}

\begin{figure}[htb]
\centerline{\includegraphics[width=0.7\columnwidth]{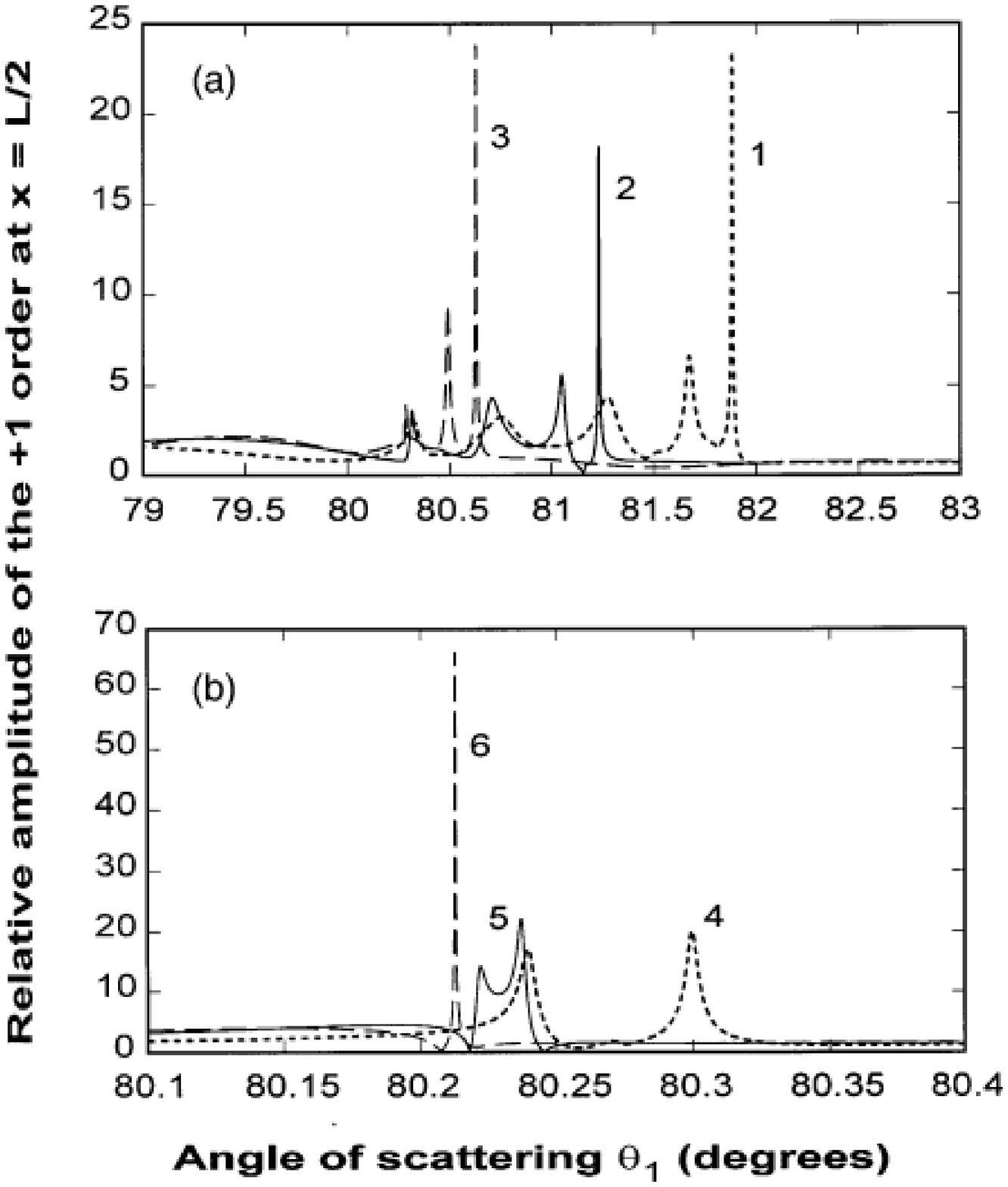}}
\caption{(a) The rigorous dependencies of the relative scattered
wave amplitude $|S_1/E_{00}|$ in the middle of the grating on angle of
scattering $\theta_1$ for the grating with $L = 10\mu$m and
(1) $\epsilon_g = 0.3$, (2) $\epsilon_g = \epsilon_{gc1} \approx 0.567$,
(3) $\epsilon_g = 0.8$. (b) The second GAS resonance.
The comparison of the rigorous dependencies of the relative
scattered wave amplitudes in the middle of two gratings of
different widths: (1) $L = 20\mu$m, $\epsilon_g = 0.8$,
(2) $L = 10\mu$m, $\epsilon_g = 0.8$ (i.e. curve 2 is the
same as curve 3 in (a)). The other
structural parameters are as before: $\epsilon = 5$,
$\theta_0 = 45^\circ$, $\lambda(\mathrm{vacuum}) = 1\mu$m.}
\end{figure}

   The presence of a critical grating amplitude $\epsilon_{gc1}$,
at which the two extreme maximums merge, producing
an extremely strong resonance (with its
subsequent rapid reduction for $\epsilon_g > \epsilon_{gc1}$, is the
general feature of GAS in different gratings. In
particular, it can be seen that generally $\epsilon_{gc1}$ depends
on grating width. However, this dependence is
weak. For example, for gratings of $L = 64.5\mu$m
(Fig. 5), $L = 20\mu$m (Fig. 6), and $L = 10\mu$m (Fig.
7a) the values of the critical grating amplitude are
0.595562, 0.5868, and 0.567, respectively. The main
features of the pattern of GAS in gratings of
smaller widths are the same as for the grating with
$L = 64.5\mu$m (compare Figs. 5--7(a)). However, the
typical height and sharpness of the resonance
maximums are significantly reduced when the
grating width is reduced---Figs. 6 and 7(a). This is
similar to the general tendency for GAS obtained
by means of the approximate theory [1]. At the
same time, it is important to understand that the
approximate theory completely fails to predict
the actual pattern of GAS at large grating amplitudes,
in particular, in the vicinity of the critical
grating amplitude.

  Note that increase of height of the resonance is
not monotonous with increasing grating amplitude%
---see curves 1--3 in Fig. 6(a). This is partly
because the grating width $L = 20\mu$m is approximately
equal to one of the optimal widths $L_\mathrm{opt}$ for
the grating amplitudes $\epsilon_g = 0.38, 0.51, 0.5868$.

   Curves 2 and 3 in Fig. 7(a) also indicate another
highly unusual and unexpected feature of scattering,
that can be revealed only using the rigorous
theory. Indeed, when the two limiting maximums
discussed above merge together (at $\epsilon_g = \epsilon_{gc1}$) and
result in a very strong and sharp resonance (dotted
curve in Fig. 5c, and curves 3 and 2 in Figs. 6(b)
and 7(a), respectively), there appears another peak
at a significantly smaller angle of scattering---see
the small sharp bump on curve 2 in Fig. 7(a) at
about $77^\circ$. When the grating amplitude is increased
($\epsilon_g > \epsilon_{gc1}$), this maximum increases substantially
and turns into a strong resonance---curve 3 in Fig.
7(a). Note that this is not the same maximum
corresponding to the described GAS resonance,
shifted towards smaller angles. Indeed, it appears
when the GAS resonance is still strong (curve 2 in
Fig. 7(a)). We will call this maximum second GAS
resonance. It exists not only in the considered
grating of $L = 10\mu$m. For example, Fig. 7(b)
presents the comparison of the second GAS resonance
for two gratings of $L = 20\mu$m (curve 1) and $L = 10\mu$m
(curve 2, which the same as curve 3 in Fig. 7(a)).
In particular, Fig. 7(b) demonstrates the general
tendency that, similar to the first GAS resonance,
the second GAS resonance becomes stronger and
sharper with increasing grating width (compare the
main maximums for curves 1 and 2 in Fig. 7(b)).
However, it is interesting that the angle of scattering,
at which the second GAS resonance occurs,
is almost independent of grating width (Fig. 7(b)).
This is again similar to the first GAS resonance---%
compare for example dotted curve in Fig. 5c
with curves 3 and 2 in Figs. 6(b) and 7(a), respectively.

\begin{figure}[htb]
\centerline{\includegraphics[width=0.7\columnwidth]{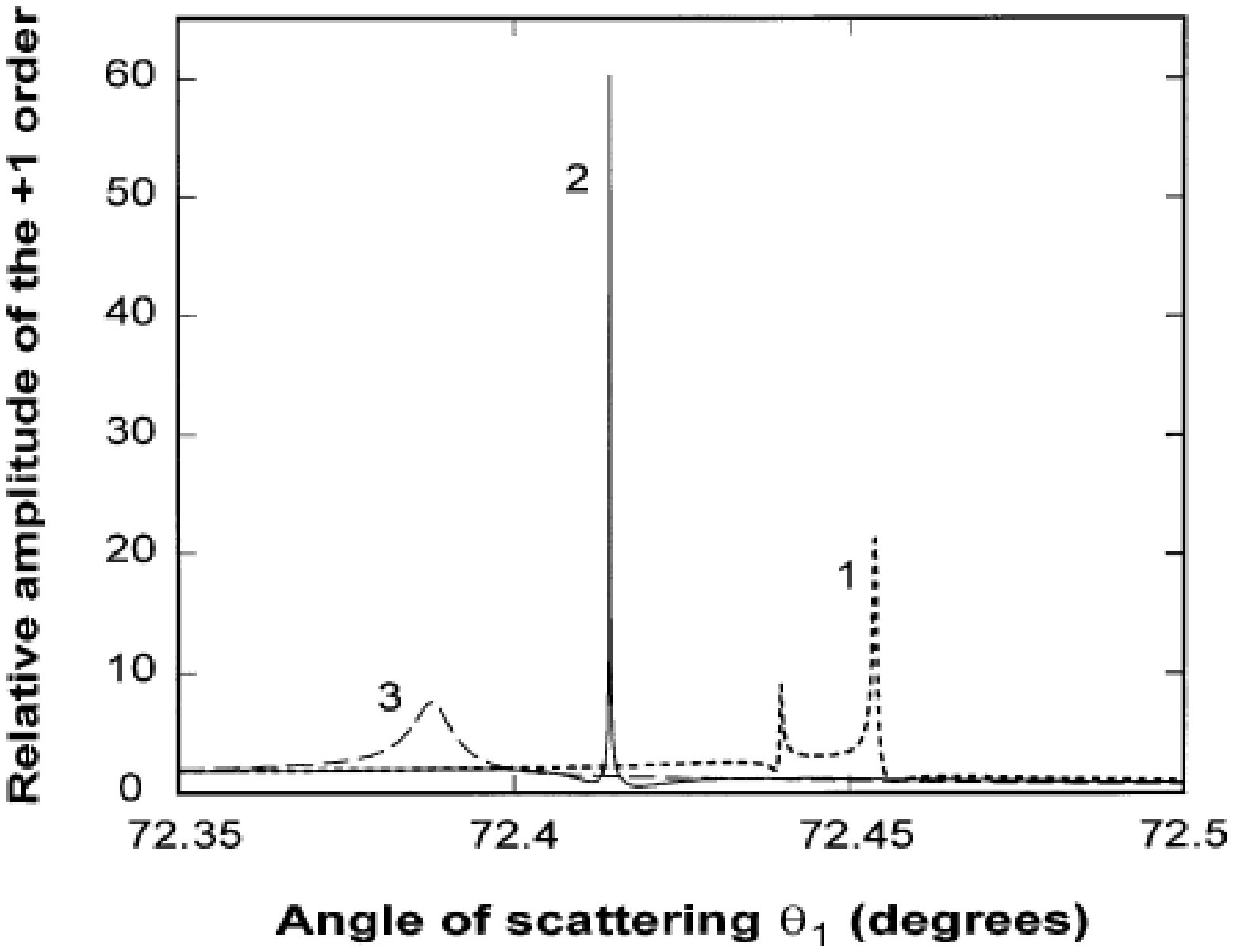}}
\caption{The second GAS resonance in the grating of $L = 20\mu$m,
$\epsilon = 5$, $\theta_0 = 45^\circ$, $\lambda(\mathrm{vacuum}) = 1\mu$m,
and (1) $\epsilon_g = 1.12$, (2)
$\epsilon_g = \epsilon_{gc2} \approx 1.1309$,
(3) $\epsilon_g = 1.14$.}
\end{figure}

    In addition, the whole pattern of scattering near
the second GAS resonance is very similar to that
demonstrated by Figs. 5(a)--(c) and 6(a) and (b)
near the first GAS resonance. For example, on the
right of the main resonant maximum the angular
dependence of the scattered wave amplitude in the
middle of the grating is fairly smooth, whereas on
the left it is characterised by a number of oscillations
with sharp maximums (Fig. 7(b)). The larger
the grating width, the larger the typical amplitude
and number of these oscillations (compare curves
1 and 2 in Fig. 7(b)). If the grating amplitude is
increased, these oscillations appear to be confined
to an angular interval between two distinct maximums
in exactly the same fashion as it was for the
first GAS resonance---see Figs. 5(a) and (b) and
6(a). One of these limiting maximums is the main
maximum corresponding to the second GAS resonance
in Fig. 7(b), and the other one is just appearing
at the angle of about $73^\circ$. Further increase
of the grating amplitude results in a significant
increase of both these limiting maximums. In the
same way as for the first GAS resonance (Figs. 5
and 6), increasing grating amplitude results in
shifting the limiting maximums (in the second
GAS resonance) towards each other with the
simultaneous reduction of the number of oscillations
between them (compare curves 1 in Figs. 8
and 7(b) with curves in Figs. 6(a) and (b)). Finally,
when the grating amplitude reaches the second
critical value $\epsilon_g = \epsilon_{gc2}$ (the index 2 stands for the
second GAS resonance), the two limiting maxi-
mums merge together producing a very strong and
sharp resonance---curve 2 in Fig. 8. Further
increase of the grating amplitude results in a rapid
decrease of the second GAS resonance (curve 3 in
Fig. 8), which is very similar to the first GAS
resonance---Fig. 5(d).

    Note also that the height of the second GAS
resonance at $\epsilon_g = \epsilon_{gc2}$ is very close to that of the
first GAS resonance at  $\epsilon_g = \epsilon_{gc1}$---compare curves
2 and 3 in Figs. 8 and 6(b), respectively. This statement
is more accurate for gratings of larger width.
It is also interesting that the value of $\epsilon_{gc2}$ is
approximately 2 times larger than $\epsilon_{gc1}$.

\begin{figure}[htb]
\centerline{\includegraphics[width=0.7\columnwidth]{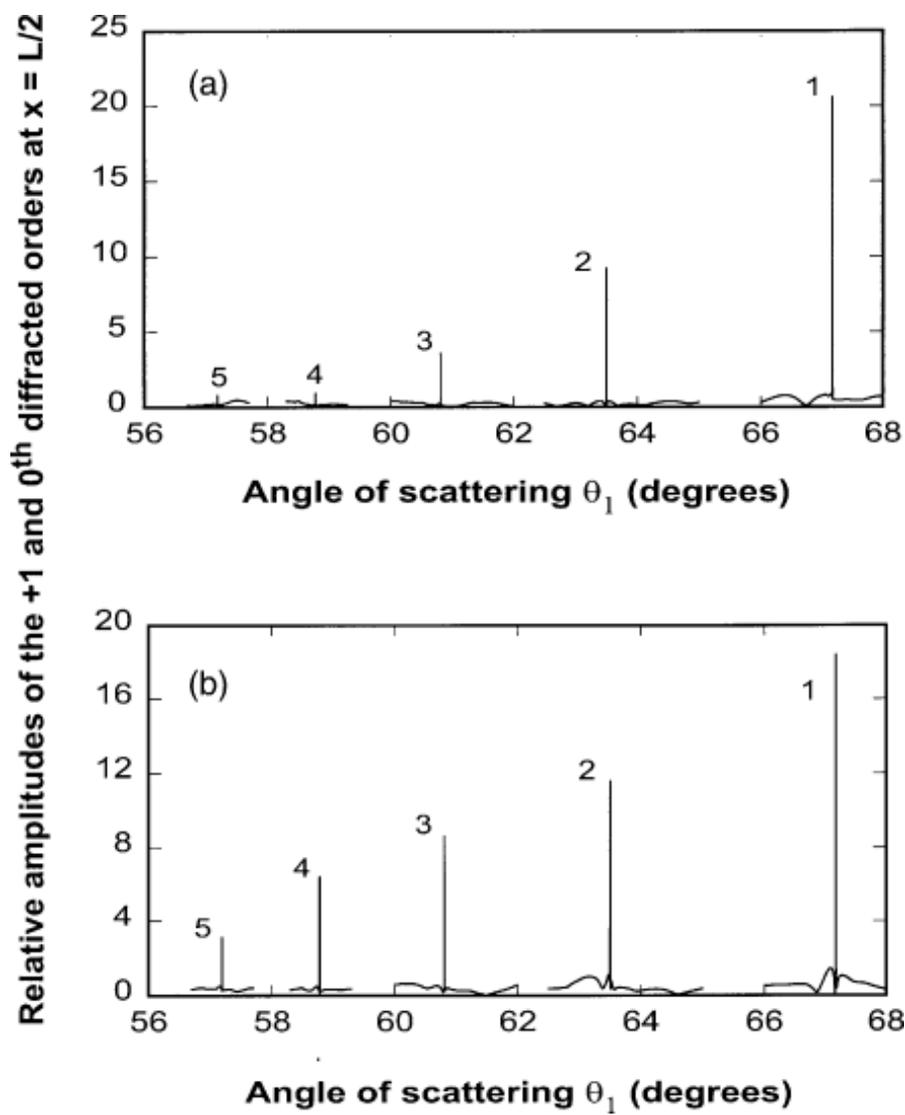}}
\caption{The rigorous dependencies of amplitudes of the scattered
(a) and incident (b) waves in the middle of the grating (i.e.
at $x = L/2$) at the critical grating amplitudes $\epsilon_{gci}$
($i = 3, 4, 5, 6,7$) for the $i$th GAS resonance:
(1) $\epsilon_g = \epsilon_{gc3} \approx 1.50988$,
(2) $\epsilon_g = \epsilon_{gc4} \approx 1.763$,
(3) $\epsilon_g = \epsilon_{gc5} \approx 1.9357$,
(4) $\epsilon_g = \epsilon_{gc6} \approx 2.055$,
(5) $\epsilon_g = \epsilon_{gc7} \approx 2.1407$.
The other structural parameters: $L = 20\mu$m, $\epsilon = 5$,
$\theta_0 = 45^\circ$, $\lambda(\mathrm{vacuum}) = 1\mu$m.}
\end{figure}

   Very strong similarities between the patterns of
scattering in the first and the second GAS resonances
suggest that when the second GAS resonance
disappears at $\epsilon_g > \epsilon_{gc2}$,
it is reasonable to
expect another resonance (third GAS resonance) to
appear at smaller angles of scattering and larger
grating amplitudes. This is indeed the case, and this
third GAS resonance is characterised by the same
pattern of scattering. That is, when the second
GAS resonance is suppressed by increasing $\epsilon_g$
above $\epsilon_{gc2}$, there appear two distinct maximums at
angles that are several degrees less than the angle at
which the second GAS resonance occurs at
$\epsilon_g = \epsilon_{gc2}$. These maximums increase with increasing
grating amplitude and shift towards each other.
Oscillations of the angular dependence of the
scattered wave amplitude occur only between these
two limiting maximums (in exactly the same way as
it was for the first and the second GAS
resonances---Figs. 5(a) and (b), 6(a), and 7(b)). Finally,
these limiting maximums merge together, producing
a very strong and sharp resonance, and this
occurs at $\epsilon_g = \epsilon_{gc3}$.
Further increase of $\epsilon_g$ results in a
rapid decrease of the third GAS resonance. After
that, the fourth GAS resonance appears at larger
grating amplitude and a smaller angle of scattering,
and so on. The corresponding resonant dependencies
of the scattered wave amplitudes in the
middle of the grating of $L = 20\mu$m and
$\epsilon_g = \epsilon_{gci}$,
where $i = 3, 4, 5, 6, 7$, are presented in Fig. 9(a).

    Fig. 9(a) clearly demonstrates that unlike the
second GAS resonance at $\epsilon_g = \epsilon_{gc2}$, that is almost
the same in height as the first GAS resonance at
$\epsilon_g = \epsilon_{gc1}$, the third, fourth, etc.
resonances at $\epsilon_g = \epsilon_{gci}$
are significantly smaller. For example, the
height of the seventh GAS resonance in the grating
of $L = 20\mu$m is only $\approx 0.76E_{00}$, where
$E_{00}$ is the
amplitude of the incident wave at the front
boundary---curve 5 in Fig. 9(a). Note also that
height of all the GAS resonances strongly depends
on grating width (unlike the values of angles and
grating amplitudes at which these resonances occur).
For example, the height of the first two GAS
resonances at $\epsilon_g = \epsilon_{gc1,2}$ is
$\approx 670E_{00}$ for $L = 64.5\mu$m,
while for $L = 20\mu$m they are $\approx 10$ times
smaller (Figs. 6(b) and 8). Similarly, the seventh
GAS resonance at $L = 20\mu$m (curve 5 in Fig. 9(a))
is also $\approx 10$ times smaller in height than the same
seventh resonance at $L = 64.5\mu$m. It is interesting
that resonance half-width is practically the same
for all the maximums in Fig. 9(a), despite the
significant decrease in their height with increasing
grating amplitude (all these maximums, including
those of curves 3 and 2 in Figs. 6(a) and 8, have the
half-widths $\approx 3 \times 10^{-4}$ deg).
At the same time, the
half-width of the resonances noticeably increases
(decreases) with decreasing (increasing) grating
width. For example, the half-width of the maximum
of curve 2 in Fig. 7(a) is $\approx 0.01^\circ$.

   Decreasing angle of incidence results in a
noticeable increase of the height and sharpness of the
predicted resonances. In addition, the resonant
angles of scattering decrease, and the critical
grating amplitude, at which the merger of two limiting
maximums occurs, increases with decreasing angle
of incidence. For example, if $L = 64.5\mu$m and
$\theta_0 = 20^\circ$ (compared to
$\theta_0 = 45^\circ$ for Figs. 2--5),
then $\epsilon_{gc1} = 1.654118$, the corresponding resonant
angle $\theta_1 \approx 73.039^\circ$, and the height of the
resonance $\approx 1300E_{00}$ (compare with
$\epsilon_{gc1} = 0.595562$,
$\theta_1 \approx 80.192^\circ$, and the resonance
height $\approx 740E_{00}$ for the dotted curve in Fig. 5c).

\begin{figure}[!t]
\centerline{\includegraphics[width=0.7\columnwidth]{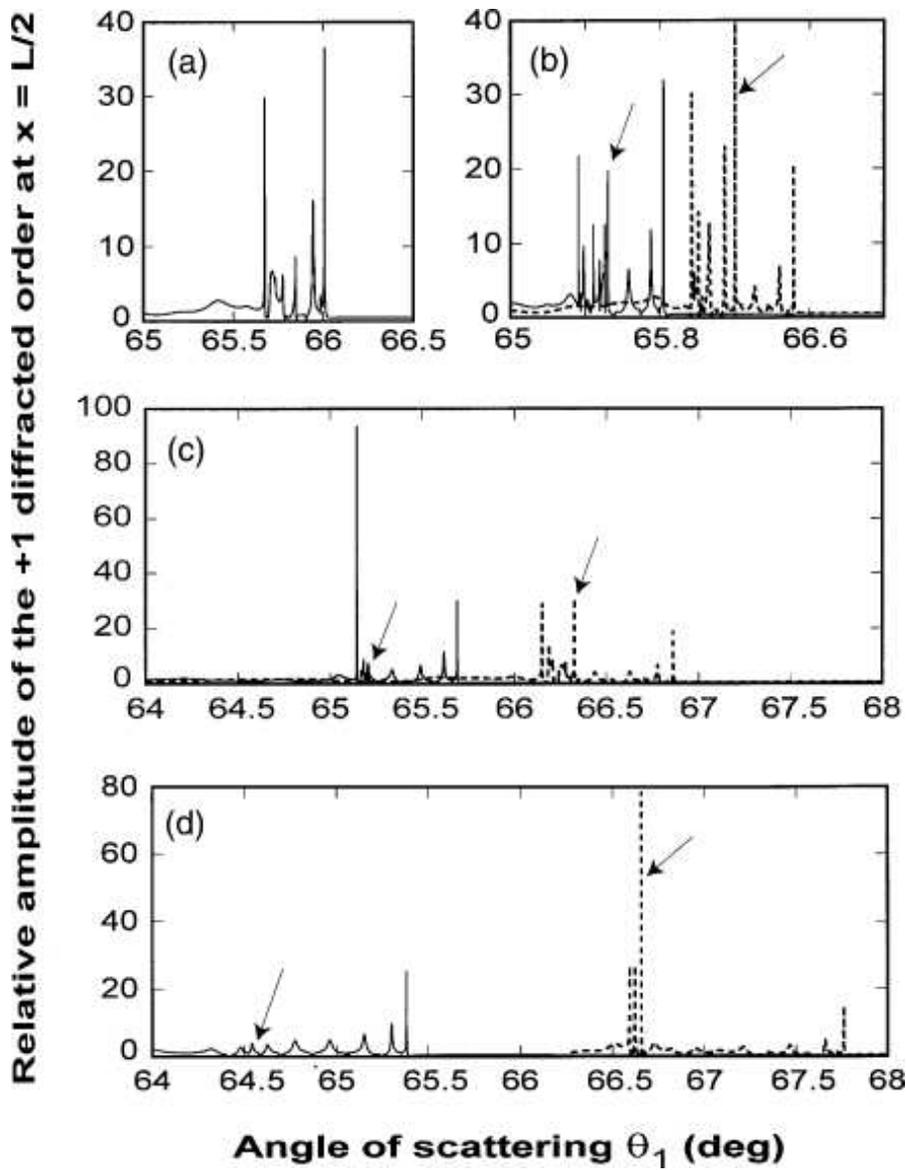}}
\caption{Small angles of incidence (near-normal incidence). The
rigorous dependencies of the relative scattered wave amplitude
in the middle of the grating of $L = 10\mu$m,
$\epsilon_g = 4$, $\epsilon = 5$,
$\lambda(\mathrm{vacuum}) = 1\mu$m.
(a) $\theta_0 = 0^\circ$, (b) $\theta_0 = 0.3^\circ$ (dotted
curve), $\theta_0 = -0.3^\circ$ (solid curve),
(c) $\theta_0 = 0.5^\circ$ (dotted curve),
$\theta_0 = -0.5^\circ$ (solid curve),
(d) $\theta_0 = 1^\circ$ (dotted curve), $\theta_0 = -1^\circ$
(solid curve). The arrows indicate the middle maximum that
splits from the right limiting maximum at $\theta_0 \ne 0$ and shifts
towards the left limiting maximum with increasing $|\theta_0|$.
}
\end{figure}

    If the angle of incidence is close to zero (almost
normal incidence onto the grating), the pattern of
scattering noticeably changes. This is illustrated by
Fig. 10, presenting the dependencies of the
amplitude of the scattered wave ($+1$ diffracted order)
on angle of scattering in the middle of the grating
of $L = 10\mu$m. If $\theta_0 = 0^\circ$,
then we obtain the pattern
with two strong limiting maximums, similar to
that in Figs. 5(a) and (b) and 6(a) (though appearing
at much larger grating amplitudes: $\epsilon_g = 4$
in Fig. 10(a)). However, if the angle of incidence is
slightly different from 0, then the right limiting
maximum in Fig. 10(a) splits into two, and there
are three distinct maximums in the pattern. The
middle (out of three) maximum is indicated by the
arrows in Figs. 10(b)--(d). As $|\theta_0|$ increases from
zero, this middle maximum splits off the right
limiting maximum in Fig. 10(a) and shifts towards
the left limiting maximum. Already when
$\theta_0 = \pm 0.3^\circ$ (Fig. 10(b)),
the pattern of scattering is
substantially different from what it was at $\theta_0 = 0^\circ$
(Fig. 10(a)). The middle maximum in Fig. 10(b) is
already more than half way through from the right
limiting maximum to the left. Note that for negative
values of $\theta_0$ the middle maximum shifts closer
to the left limiting maximum than for the same
positive values (compare solid and dotted curves in
Fig. 10(b)). If $|\theta_0|$ is increased further, the middle
maximum may vary strongly in height and
sharpness (Figs. 10(b--d)). Finally, the middle and
the left maximums for $\theta_0 < 0$ (solid curves) merge
together, producing a very high resonance, and
further increase of $|\theta_0|$ quickly results in
annihilation of both the maximums (solid curve in Fig.
10(d)). A similar pattern is observed when the
angle $\theta_0$ in increased from zero---see the dotted
curves in Figs. 10(b)--(d). However, in this case the
merger of the left and the middle maximums occur
at larger values of $\theta_0$---compare Figs. 10(c) and
(d). The right limiting maximum tends to decrease
with increasing $\theta_0$. Therefore, it may have
noticeable height only at almost normal incidence---%
typically for $\theta_0 \le 5^\circ$.

   From here, we can see the relationship between
the pattern of scattering presented by Figs. 5--7a for
large incidence angles and the pattern shown by
Fig. 10. As mentioned above, decreasing (increasing)
angle of incidence results in increasing (decreasing)
values of $\epsilon_{gc1}$. Therefore, instead of fixing
$\theta_0$ and finding $\epsilon_{gc1}$,
we can fix $\epsilon_g$ and choose the
critical angle of incidence $\theta_{0c}$ corresponding to
the merger of the maximums. This suggests that
the middle and the left limiting maximums in
Figs. 10(b)--(d) correspond to the two limiting
maximums considered in Figs. 5--7a. The right
limiting maximum in Figs. 10(b)--(d) could not be
seen in Figs. 5--7a since it is negligible at large
angles of incidence.

   Note also that according to the tendency
mentioned above, resonant angles of scattering
decrease with decreasing angle of incidence. This is
also in obvious agreement with Figs. 10(a)--(d).

\begin{figure}[htb]
\centerline{\includegraphics[width=0.7\columnwidth]{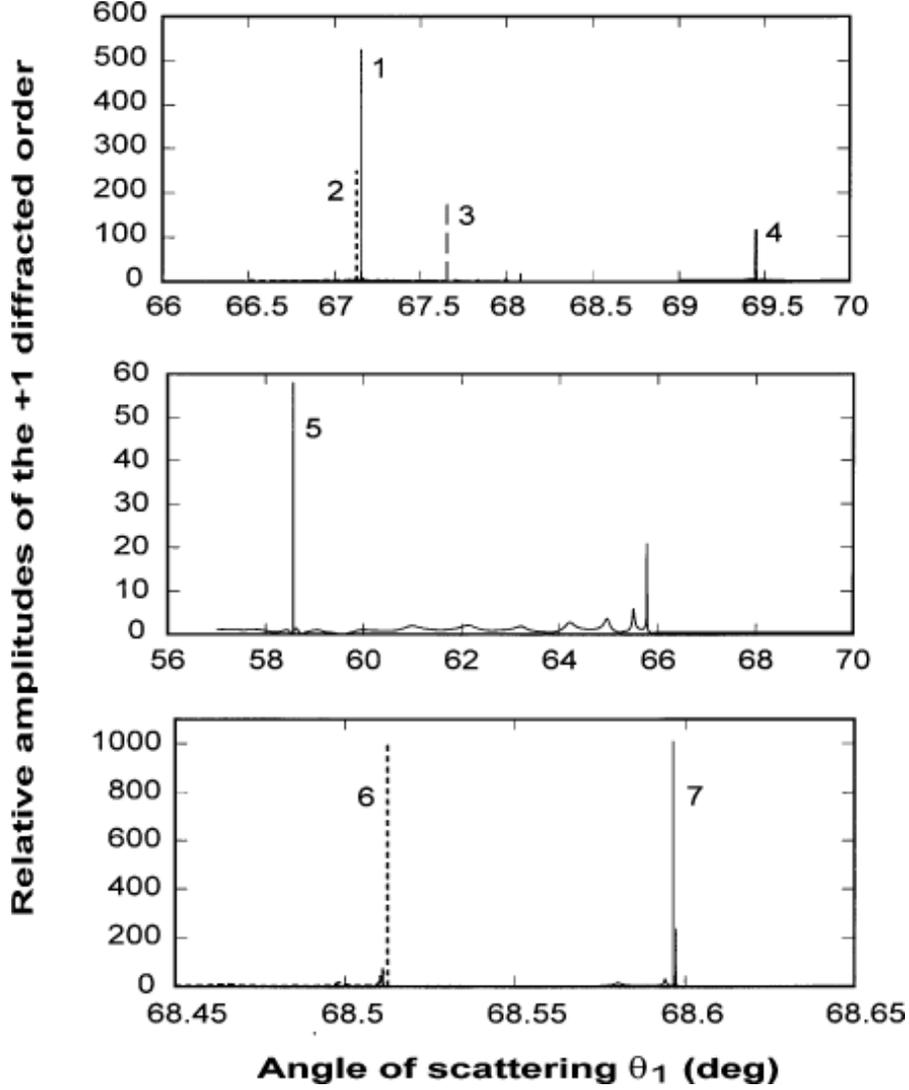}}
\caption{The rigorous dependencies of the scattered wave
amplitude in the middle of the grating when the left and the middle
maximums merge, i.e. at $\epsilon_ge = \epsilon_{gc1}$.
The grating width is optimised for the strongest merged
resonance in each grating. (1)
$L = 10\mu$m, $\epsilon_ge = \epsilon_{gc1} \approx 4.2439906$,
$\theta_0 = 1^\circ$, resonance halfwidth
$\delta\theta \approx 7\times 10^{-6}$ deg, (2) $L = 9.93\mu$m,
$\epsilon_ge = \epsilon_{gc1} \approx 3.53973$,
$\theta_0 = 3^\circ$, $\delta\theta \approx 4\times 10^{-5}$ deg,
(3) $L = 10.05\mu$m, $\epsilon_ge = \epsilon_{gc1} \approx 3.1133$,
$\theta_0 = 5^\circ$, $\delta\theta \approx 6\times 10^{-5}$ deg,
(4) $L = 9.97\mu$m, $\epsilon_ge = \epsilon_{gc1} \approx 2.4325$,
$\theta_0 = 10^\circ$, $\delta\theta \approx 1.5\times 10^{-4}$ deg,
(5) $L = 10\mu$m, $\epsilon_ge = \epsilon_{gc1} \approx 2.3624$,
$\theta_0 = -3^\circ$, $\delta\theta \approx 4\times 10^{-5}$ deg,
(6) $L = 10.05\mu$m, $\epsilon_ge = \epsilon_{gc1} \approx 5.1491$,
$\theta_0 = 0$, $\delta\theta \approx 9\times 10^{-7}$ deg,
(7) $L = 10\mu$m, $\epsilon_ge = \epsilon_{gc1} \approx 5.1823$,
$\theta_0 = 0$, $\delta\theta \approx 3\times 10^{-7}$ deg.}
\end{figure}

   As mentioned above, when the middle and the
left limiting maximums in Figs. 10(b)--(d) merge,
they produce extremely strong and sharp resonances
(similar to those in Figs. 5c, 6b, 7a, 8 and
9). This merger can be achieved by choosing the
right angle of incidence and/or grating amplitude.
The resultant typical resonance maximums are
presented in Fig. 11. The corresponding structural
parameters and half-widths of the maximums are
presented in the figure caption. Note that for all
resonances in Figs. 11 the grating width has been
optimised for maximal height of the main maximums.
This is the reason for slight variations of
the grating widths of the considered gratings.

\begin{figure}[!b]
\centerline{\includegraphics[width=0.7\columnwidth]{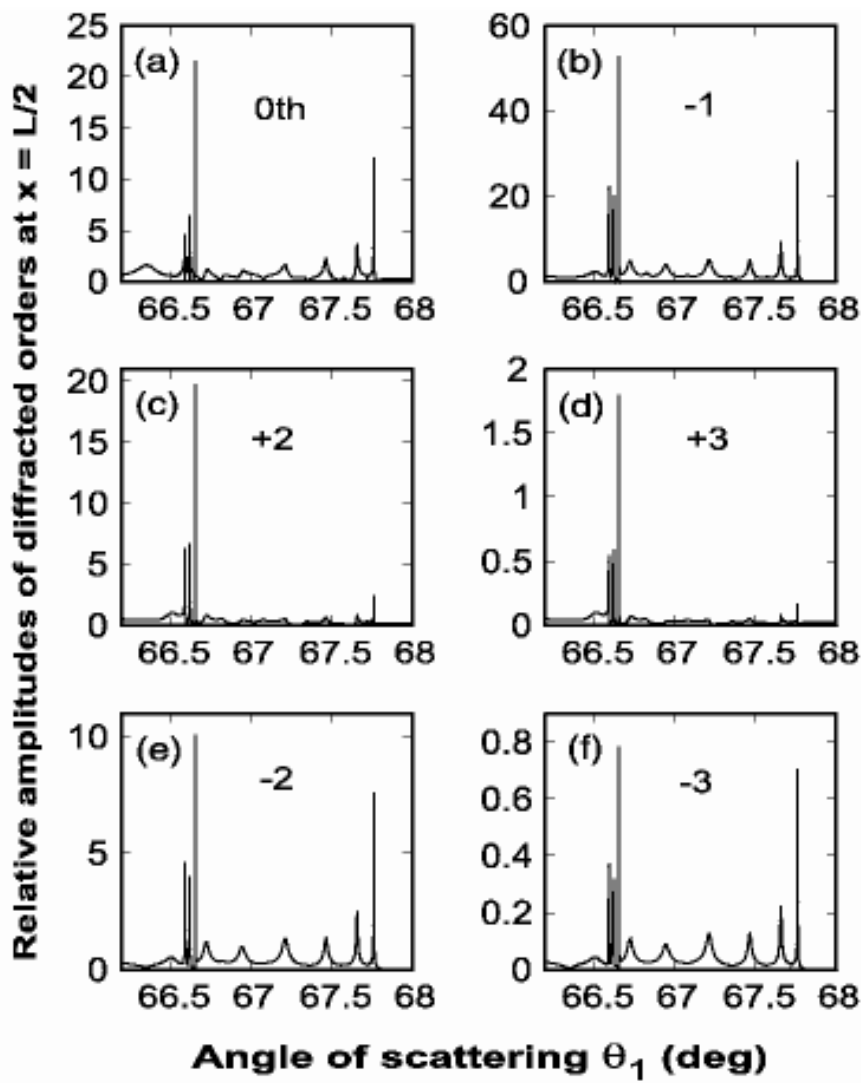}}
\caption{The rigorous dependencies of the relative amplitudes
of the six diffracted orders (having significant amplitudes) on
angle of scattering $\theta_1$ for the structure corresponding to the
dotted curve in Fig. 10(d). The number of the particular
diffracted order is indicated in each of the subplots.}
\end{figure}

   It can be seen that decreasing angle of incidence
results in a substantial increase of height of the
merged maximums (compare curves 1--5 in
Fig. 11). The smaller maximums of curves 5 and 3
(for curve 3 it occurs at $\approx 68.1^\circ$---Fig. 11(a))
correspond to the right limiting maximums in
Figs. 10(b)--(d). For positive $\theta_0$, this maximum
completely disappears at $\theta_0 > 10^\circ$, whereas for
negative $\theta_0$ it is much more noticeable even for
non-normal incidence (though still relatively
small).

    At normal incidence, there is no middle maximum
(Fig. 10(a)), and the merger occurs between
the two limiting maximums. The resultant merged
resonance is especially strong---curve 6 in Fig. 11.
In this case, further increase of the grating
amplitude results in an oscillatory behaviour of the
resonance. For example, increasing grating
amplitude beyond the value corresponding to curve 6
($\epsilon_g = 5.1491$) results in a significant reduction of
the resonance (to $\approx 300E_{00}$), and then in increasing
it back to $\approx 1090E_{00}$ (at
$\epsilon_g = 5.1823$---see curve 7 in
Fig. 11). Further increase of the grating amplitude
results in other maximums that exceed several
thousands of $E_{00}$.

   It is obvious that the predicted extremely large
amplitudes of the $+1$ diffracted orders, together
with the large grating amplitude, must lead to very
significant amplitudes of other diffracted orders in
the Floquet expansion (2). This is demonstrated by
Fig. 12, where amplitudes of several diffracted
orders, other than the $+1$ order, are presented for
the structure corresponding to the dotted curve in
Fig. 10(d).

   It can be seen that the amplitude of the incident
wave (0th diffracted order) approximately follows
the $+1$ diffracted order---compare the dotted curve
in Fig. 10(d) and the curve in Fig. 12(a). This is
similar to what has been predicted previously for
GAS by means of the approximate and rigorous
theories (see [1] and Fig. 2(b)). In particular, both
the scattered and incident wave amplitudes experience
a strong resonant increase in the middle of
the grating. This is expected, since the extremely
large scattered wave amplitude inside the grating
must result in substantial re-scattering, as a result
of which the incident wave must also have a large
amplitude (note however that the energy conservation
requires the incident wave amplitude at the
rear grating boundary to be $\le E_{00}$). The $+2$
diffracted order is also expected to be large, since it is
directly coupled to the resonantly strong $+1$
diffracted order (Fig. 12c). However, what is really
surprising is that the $-1$ diffracted order is
significantly stronger than the 0th and $+2$ orders. This is
unexpected, since the $-1$ order is not coupled directly
to the resonantly strong scattered wave, and
one could think that it should be weaker than the
0th and $+2$ orders. The rigorous analysis demonstrates
that this expectation is not correct, and the
amplitude of the $-1$ order is not only larger than
those of the 0th and $+2$ orders, but is very close to
the amplitude of the $+1$ order (i.e. scattered wave).
The same situation is with the $-2$ order that has (in
the middle maximum) the amplitude only $\approx 2$ times
less than that of the $+2$ order, though it can be
expected to be significantly smaller.

\begin{figure}[!t]
\centerline{\includegraphics[width=0.7\columnwidth]{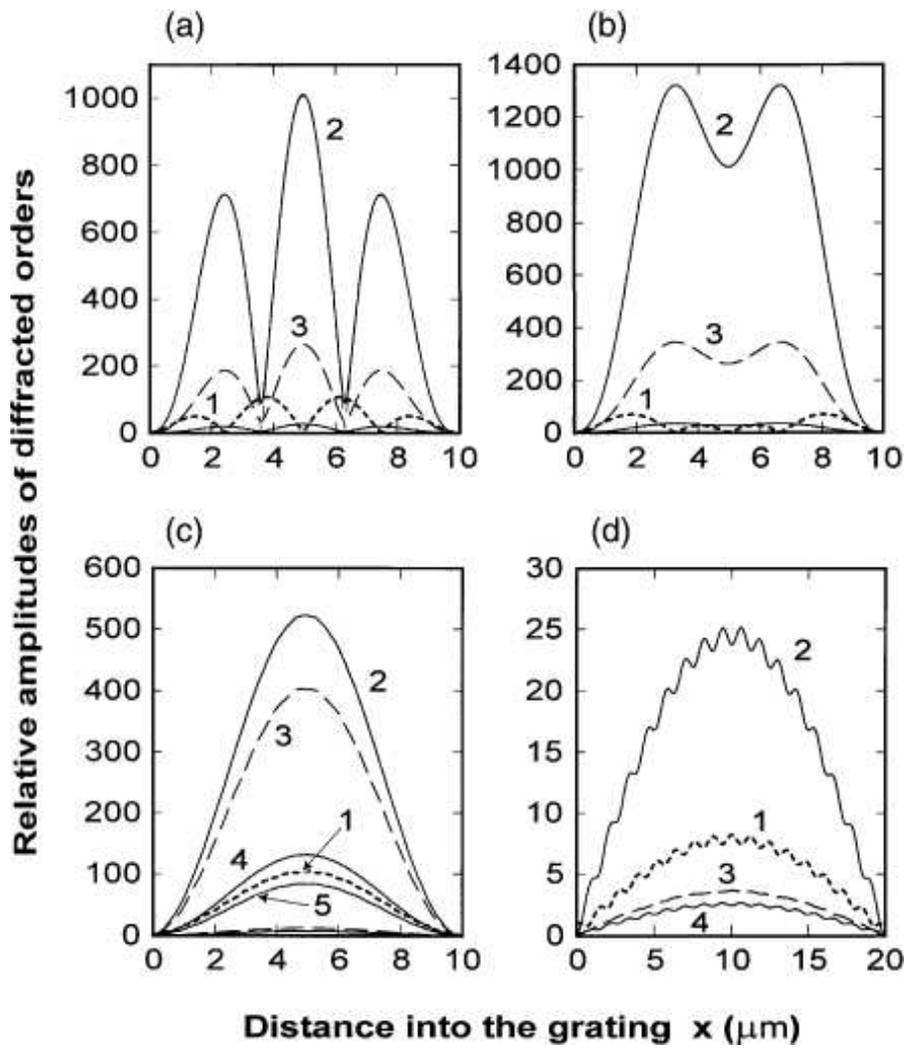}}
\caption{The rigorous $x$-dependencies of the relative amplitudes
of several diffracted orders with noticeable amplitude inside the
grating at the resonance angles of scattering $\theta_{1\mathrm{r}}$.
The subplots
(a)--(d) correspond to the resonances of curves 6, 7, 1 in Fig. 12,
and curve 3 in Fig. 6, respectively. (a) 0th order (curve 1), $+1$
and $-1$ orders (curve 2), $+2$ and $-2$ orders (curve 3), $+3$ and $-3$
diffracted orders (lower solid curve);
$\theta_1 = \theta_{1\mathrm{r}} \approx 68.51250565^\circ$.
(b) 0th order (curve 1), $+1$ and $-1$ orders (curve 2), $+2$ and $-2$
orders (curve 3), $+3$ and $-3$ diffracted orders (lower solid curve);
$\theta_1 = \theta_{1\mathrm{r}} \approx 68.59630026^\circ$.
(c) 0th order (curve 1), $+1$ order
(curve 2), $-1$ order (curve 3), $+2$ order (curve 4), $-2$ order
(curve 5), $+3$ and $-3$ orders (lower dashed and solid curves,
respectively); $\theta_1 = \theta_{1\mathrm{r}} \approx 67.1545608^\circ$.
(d) 0th order (curve 1),
$+1$ order (curve 2), $-1$ order (curve 3), $+2$ order (curve 4);
$\theta_1 = \theta_{1\mathrm{r}} \approx 80.6295^\circ$.}
\end{figure}

    This effect is even more obvious if we calculate
amplitudes of the diffracted orders for the structures
corresponding to curves 6 and 7 in Fig. 11. In
this case, the angular dependencies of the amplitudes
of the $-1$ and $+1$ orders are practically
indistinguishable. The same is true for the $-2$ and $+2$
orders, $-3$ and $+3$ orders, etc. In addition, all these
dependencies almost exactly reproduce the shape
of curves 6 and 7 in Fig. 11 (though of different
maximal heights). Therefore, presentation of these
dependencies here is not worthwhile. Instead, Fig.
13 presents the typical $x$-dependencies of amplitudes
of several diffracted orders in the gratings of
widths $L \approx 10\mu$m (Fig. 13a--c) and $L = 20\mu$m
(Fig. 13(d)). The subplots in Fig. 13 correspond to
the maximums of curve 6 in Fig. 11 (Fig. 13(a)),
curve 7 in Fig. 11 (Fig. 13(b)), curve 1 in Fig. 11
(Fig. 13c), and curve 3 in Fig. 6(a) (Fig. 13(d)). In
accordance with the mentioned above, the $x$-dependencies
of the amplitudes of the $+1$ and $-1$
diffracted orders in Figs. 13(a) and (b) are
practically indistinguishable and are both represented by
curves 2. The match of these dependencies is better
for Fig. 13(b), while in Fig. 13(a) there are minor
differences near the two minimums of curve 2 in
the grating (not shown in the figure). Similarly,
each of curves 3 in Figs. 13(a) and (b) simultaneously
represent the amplitudes of the $+2$ and $-2$
orders, and the lower solid curves in both the
figures correspond to the $+3$ and $-3$ orders. This
result is completely unexpected and surprising.
Indeed, despite the obvious non-symmetry of the
structure (Fig. 1) with respect to scattering into,
for example, $+1$ and $-1$ orders (e.g., the Bragg
condition is satisfied only for the $+1$ order), the
amplitudes of these orders (and all other $+n$ and
$-n$ orders) are practically indistinguishable in Figs.
13(a) and (b).

   This however is not the case for less sharp and
high resonances at larger angles of incidence---%
Figs. 13(c) and (d). In these figures, the amplitudes
of the $+1$ and $-1$ orders (as well as those of the $+2$
and $-2$ orders, etc.) are noticeably different.
Nevertheless, the stronger the resonance, the closer the
amplitude of the $-1$ order to that of the $+1$ order
(compare Figs. 13(c) and (d)). It is worth noting
that strong resonances, as in Fig. 11, are highly
sensitive to grating width. For example, if for
curve 6 in Fig. 11 the grating width is changed
from 10.02 to $10.1\mu$m, then the maximum of curve
6 in Fig. 11 reduces to $\approx 420E_{00}$ (i.e. more than two
times compared to what it is in Fig. 11).

   The small oscillations displayed by all the
curves in Fig. 13(d) are also typical for other
subplots in Fig. 13. However, due to much smaller
scale, these oscillations are hardly seen on curves
in Fig. 13a--c.

   It is also important to realise that the presented
results are not only relevant to the particular
structural parameters considered above. It can be
shown that the scaling procedure described in [12]
is readily applicable to the considered gratings.
For example, if the mean permittivity together
with the grating amplitude are increased $\alpha$ times,
then decreasing grating width $\alpha^{1/2}$ times must
result in exactly the same results as for the gratings
discussed above.

\section{Eigenmodes of a slanted grating}

   The results obtained in Sections 3 and 4
demonstrate an extremely complex pattern of
scattering, involving a number of strong resonances
associated with strong increase of amplitudes of
several diffracted orders inside the grating. It is
obvious that a comprehensive physical explanation
of all the predicted effects and resonant behaviour
is hardly feasible at this stage. However, it
is rather clear that diffractional divergence of the
scattered wave, that has been used for the explanation
of wave effects in the geometry of EAS [3--8],
can hardly be used for the explanation of the
predicted resonances. This is because if the grating
amplitude is large, these resonances occur at angles
of scattering significantly different from $\pi/2$. For
example, at almost normal incidence, strong resonances
have been found at angles of scattering $\theta_1$
between $40^\circ$ and $50^\circ$. It is hardly possible to expect
that diffractional divergence of the $+1$ order (or
any other diffraction order) can play a significant
role at such angles of propagation with respect to
the grating boundaries (see also [3--8]). Moreover,
though the resonances were frequently referred to
as GAS resonances in Section 4, one should not be
deceived by this terminology, since in some cases
the scattered wave no longer propagates at a
grazing angle with respect to the grating
boundaries (see also Figs. 9--12).

   The explanation of the observed extremely
strong resonances can be understood from Fig. 13.
This figure demonstrates that in the strongest
resonances, the amplitudes of the diffracted orders
are resonantly large only inside the grating,
whereas at its boundaries they are close to zero.
On the other hand, any sufficiently strong resonance
is associated with generation of some kind
of eigen oscillations or eigenmodes in the structure.
Therefore, the discovered resonances must be
related to the resonant generation of a special new
type of grating eigenmodes by an incident wave
(the grating eigenmodes are coupled to the incident
wave). Actually, these modes are not true
structural eigenmodes, since if they were, they
would have not been coupled to a incident wave.
Due to the presence of this weak coupling, grating
eigenmodes weakly leak from the grating. However,
since the predicted resonances are extremely
high (up to hundreds or even thousands of the
amplitude of the incident wave), the leakage must
be weak. In the case of a high resonance, the field
in the grating represents, to a high degree of
accuracy, the field in the corresponding grating
eigenmode. Thus the field distribution presented in
Fig. 13 is actually the field distribution in the
grating eigenmodes. This is because the perturbation
effect of the relatively weak incident wave
(with $E_{00} \ll S_1|_{x=L/2}$) on the field distribution inside
the grating is negligible.

   It is important to recall that the considered
gratings are not associated with any conventional
guiding effect, since the mean dielectric permittivity
$\epsilon$ is assumed to be the same inside and outside
the grating (see Section 2 and Fig. 1). The guiding
effect on the eigenmodes is imposed only by the
grating. As a result, a grating of $10\mu$m width can
guide a wave with the amplitude in the middle of
the grating, that is thousands of times larger than
at the grating boundaries. It is also interesting that
all the gratings considered in this paper are oblique
(slanted) gratings, which makes them asymmetric
from the view-point of a mode propagating along
the grating. Nevertheless, the field distribution in
the modes corresponding to strong resonances is
always practically symmetric with respect to the
middle of the grating (Figs. 13(a)--(c)).

    The discovered eigenmodes are formed by the
interacting diffracted orders in the grating. This
interaction occurs so that amplitudes of the
diffracted orders decrease to about zero as they
propagate towards a grating boundary. For example,
the diffracted orders whose wave vectors
point towards the rear grating boundary increase
in amplitude (gaining energy from the diffracted
orders ``travelling'' in the opposite direction) in the
first half of the grating (from $x = 0$ to $x = L/2$),
and then lose their energy due to the same interaction
in the second half of the grating. The opposite
occurs for diffraction orders with the wave
vectors pointing towards the front boundary.

   Since we can consider a wave incident on the
grating from its either side, the discovered
eigenmodes can equally propagate in both directions
along the grating, i.e. in the positive and negative
directions of the $y$-axis in Fig. 1.

   As can be seen from Fig. 13, the structure of the
grating eigenmodes can be quite different, de-
pending on structural parameters and angles of
propagation of diffraction orders (i.e. mode type).
These modes are significantly more complex than
the conventional modes of a slab waveguide. This
has also been demonstrated by the consideration
of grating eigenmodes in the presence of the
conventional guiding effect, i.e. in a guiding slab with
a modulated dielectric permittivity [13]. In this
case, a number of new grating eigenmodes are
predicted, that are strongly different from the
conventional guided modes in a slab (e.g., grating
eigenmodes in a slab cannot exist in the absence of
the grating) [13].

   Note that without the consideration of the
grating eigenmodes the discussed strong resonances,
like those in Figs. 11(a)--(c), are practically
useless. This is because they are very difficult to
achieve in practice due to extremely large relaxation
times, and thus impractically large apertures
of the incident bean that would be required for the
steady-state to be achieved. However, the existence
of the grating eigenmodes radically changes the
situation. The resonances are used only for the
determination of the field structure of these modes
that can be generated by other means, similar to
those used for generation of the conventional slab
modes.

   Note also that especially strong resonances (and
the described grating eigenmodes) are often
obtained for very large grating amplitudes (e.g.,
$\epsilon_g \approx 3$ to 5). Recalling that, according to Eq. (1),
the actual amplitude of modulation of the mean
permittivity in the grating is equal to
$2\mathrm{Re}(\epsilon_g)$, we
can see that $\epsilon_g \approx 5$ corresponds to the modulation
amplitude $2\epsilon_g \approx 10$. This is 2 times larger than the
mean permittivity in the structure. This means that
the permittivity in the grating must vary from
positive to negative values. On the other hand,
negative permittivity can be obtained in metals,
ionic crystals (between the frequencies of transverse
and longitudinal optical phonons), or near
any sufficiently strong material resonance. In this
case, dissipation in the medium is inevitable. The
accurate analysis of the grating eigenmodes and
the associated resonances in the presence of
dissipation is beyond the scope of this paper. However,
it can be noted that small dissipation could, for
example, be compensated by appropriate gain in
the material of the grating.

\section{Conclusions}

   In this paper, the detailed rigorous analysis of
GAS of bulk TE electromagnetic waves in slanted
uniform holographic gratings has been carried out.
This analysis has demonstrated high accuracy of
the previously developed approximate theory for
gratings with small amplitudes (up to $\approx 1$\% of the
mean permittivity), especially for near-normal
incidence. Even if the discrepancies between the
approximate and rigorous theories are noticeable
(at grating amplitudes that are less than $\approx 10$\% of
the mean permittivity), they are mainly restricted
to variations of resonant angles, but not the shape
of the curves and the height of the GAS resonances.

   On the other hand, a highly unusual, unexpected,
and complex pattern of resonant behaviour
of several diffracted orders in gratings has
been discovered at large grating amplitudes
(greater than $\approx 10$\% of the mean permittivity). The
resonant angles in this case lie within the large
range and do not necessarily correspond to the
geometry of GAS (where the $+1$ diffracted order
propagates at a grazing angle with respect to the
grating boundaries). Even in relatively narrow
gratings (of $\approx 10\mu$m) these new resonances can be
extremely strong (up to thousands of the amplitude
of the incident wave at the front boundary).
Increasing grating amplitude generally results in a
rapid increase of the corresponding resonances.

    Though the analysis was carried out only for
sinusoidal gratings, it is highly likely that similar
resonances occur in gratings with arbitrary profiles,
as well as in two-dimensional and three-dimensional
grating with large amplitude (i.e.
photonic crystals). The analysis of such structures
is currently being carried out.

    Physical explanation of the predicted resonant
behaviour of waves in the grating has been linked
to the generation of a special new type of grating
eigenmodes. These modes are guided by a slanted
grating with large amplitude, like modes guided by
a slab. However, grating eigenmodes are shown to
have much more complex structure with several
diffracted orders involved. The field distribution in
such modes has been investigated and discussed.

   It is obvious that for the predicted resonances
to be achieved experimentally, the corresponding
time of relaxation to the steady-state regime of
scattering must be reasonably small. Otherwise,
the aperture of the incident beam that could be
required for achieving the steady-state regime
would be impractically large [4,5,14]. The determination
of which of the predicted resonances can
reasonably be achieved in practice, together with
the analysis of non-steady-state scattering, can be
carried out by methods developed for non-steady-state EAS in [14].

   On the other hand, the main importance of the
discovery of the strongest predicted resonances
(that are obviously not achievable in practice due
to large relaxation times) is in the two main results.
First, they demonstrate radically new, previously
unseen resonant effects in slanted gratings. Second,
the existence of grating eigenmodes and their field
structure are derived from the consideration of
these resonances. In the end, the discovered
eigenmodes can well be generated by means other
than the resonant scattering of the incident wave in
the grating (for example, by means similar to those
used for generation of slab modes).

\section*{Acknowledgements}

  The authors gratefully acknowledge financial
support for this research from the Queensland
University of Technology.

\section*{References}

\begin{enumerate}
\item D.K. Gramotnev, Opt. Quant. Electron. 33 (2001) 253.
\item S. Kishino, A. Noda, K. Kohra, J. Phys. Soc. Jpn. 33
     (1972) 158.
\item M.P. Bakhturin, L.A. Chernozatonskii, D.K. Gramotnev,
     Appl. Opt. 34 (1995) 2692.
\item D.K. Gramotnev, Phys. Lett. A 200 (1995) 184.
\item D.K. Gramotnev, J. Phys. D 30 (1997) 2056.
\item D.K. Gramotnev, Opt. Lett. 22 (1997) 1053.
\item D.K. Gramotnev, D.F.P. Pile, Phys. Lett. A 253 (1999)
     309.
\item D.K. Gramotnev, D.F.P. Pile, Opt. Quant. Electron. 32
     (2000) 1097.
\item T.A. Nieminen, D.K. Gramotnev, Opt. Commun. 189
     (2001) 175.
\item M.G. Moharam, E.B. Grann, D.A. Pommet, T.K. Gaylord,
    J. Opt. Soc. Am. A 12 (1995) 1068.
\item M.G. Moharam, D.A. Pommet, E.B. Grann, T.K. Gaylord,
     J. Opt. Soc. Am. A 12 (1995) 1077.
\item D.K. Gramotnev, T.A. Nieminen, T.A. Hopper, J. Mod.
     Opt. 49 (2002) 1567.
\item D.K. Gramotnev, S.J. Goodman, T.A. Nieminen, in press.
\item D.K. Gramotnev, T.A. Nieminen, Opt. Express 10 (2002)
     268.
\end{enumerate}

\end{document}